\documentclass[twocolumn]{openjournal}[revtex]

\usepackage[breaklinks,colorlinks,citecolor=blue,linkcolor=blue,urlcolor=blue]{hyperref}
\usepackage{graphicx}
\usepackage{amsmath}

\begin{document}

\title{Bayesian Cosmic Void Finding with Graph Flows}

\author{Leander Thiele}
\email{leander.thiele@ipmu.jp}
\affiliation{Center for Data-Driven Discovery, Kavli IPMU (WPI), UTIAS, The University of Tokyo, Kashiwa, Chiba 277-8583, Japan}
\affiliation{Kavli IPMU (WPI), UTIAS, The University of Tokyo, 5-1-5 Kashiwanoha, Kashiwa, Chiba 277-8583, Japan}

\begin{abstract}
  Cosmic voids contain higher-order cosmological information and are of interest for astroparticle physics.
  Finding genuine matter underdensities in sparse galaxy surveys is, however, an underconstrained problem.
  Traditional void finding algorithms produce deterministic void catalogs, neglecting the probabilistic nature of the problem.
  We present a method to sample from the stochastic mapping from galaxy catalogs to arbitrary void definitions.
  Our algorithm uses a deep graph neural network to evolve ``test particles'' according to a flow-matching objective.
  We demonstrate the method in a simplified example setting but outline steps to generalize it towards practically usable void finders.
  Trained on a deterministic teacher, the model performs well but has considerable stochasticity which we interpret as regularization.
  Cosmological information in the predicted void catalogs outperforms the teacher.
  On the one hand, our method can cheaply emulate existing void finders with apparently useful regularization.
  More importantly, it also allows us to find the Bayes-optimal mapping between observed galaxies and any void definition.
  This includes definitions operating at the level of simulated matter density and velocity fields.
\end{abstract}

\maketitle

\section{Introduction}\label{sec:intro}

The late-time density field has grown gravitationally out of tiny initial perturbations into the intricate cosmic web.
Extracting information about fundamental physics from this large-scale structure is one of the primary goals in current observational cosmology.
Due to nonlinear growth, an optimal analysis method is difficult to conceive.
Thus, various summary statistics have been developed to probe different aspects of the density field.
Volume-wise the large-scale structure is dominated by underdense regions: cosmic voids.
Voids have been recognized as holding unique cosmological information~\citep[see e.g.,][]{Pisani2019,Moresco2022,Schuster2023,Schuster2024,Schuster2025,Contarini2026}.
Finding voids has been an active area of research, an attempt at a comprehensive list of void finders being~\cite{El-Ad1997,Hoyle2002,Padilla2005,Neyrinck2008,Lavaux2010,Sutter2015,Elyiv2015,Ruiz2015,Zhao2016,Banerjee2016,Nadathur2019,Ruiz2019,Douglass2022,Paz2023,Monllor-Berbegal2025,Ghafour2025,Sartori2026}.
It has been recognized that due to the multi-scale nature of the cosmic web the definition of a void is not unique, and indeed an ``optimal'' definition depends on the science case~\citep{Colberg2008,Cautun2018,Paillas2019}.
Additionally, void finders can be modified through postprocessing of the output catalogs~\citep{Ronconi2017,Verza2025}.

The void finding literature has strongly focused so far on addressing the problem with a deterministic approach, neglecting the fact that it naturally maps to probabilistic methods.
Restricting the discussion to the case of galaxy redshift surveys, the observables are sparse and biased tracers of the density field.
Thus, finding the underdense zones is an underconstrained problem.
What seems more meaningful to consider is the probability distribution over void catalogs conditioned on galaxy catalogs
\begin{equation}
  P(\{v_\alpha\}_{\alpha=1 \ldots N_v} | \{g_\beta\}_{\beta=1 \ldots N_g})\,,
  \label{eq:catalogp}
\end{equation}
where the $v_\alpha$ and $g_\beta$ are tuples containing void and galaxy properties, respectively.
Traditional void finders adopt deterministic definitions of $\{v_\alpha\}$ such that this distribution collapses to a delta function.
There are some algorithms computing confidence scores for individual voids, but due to void exclusion and similar effects such factorization of the distribution is a poor approximation: one needs to take correlations between voids into account.

\begin{figure}
  \includegraphics[width=\linewidth]{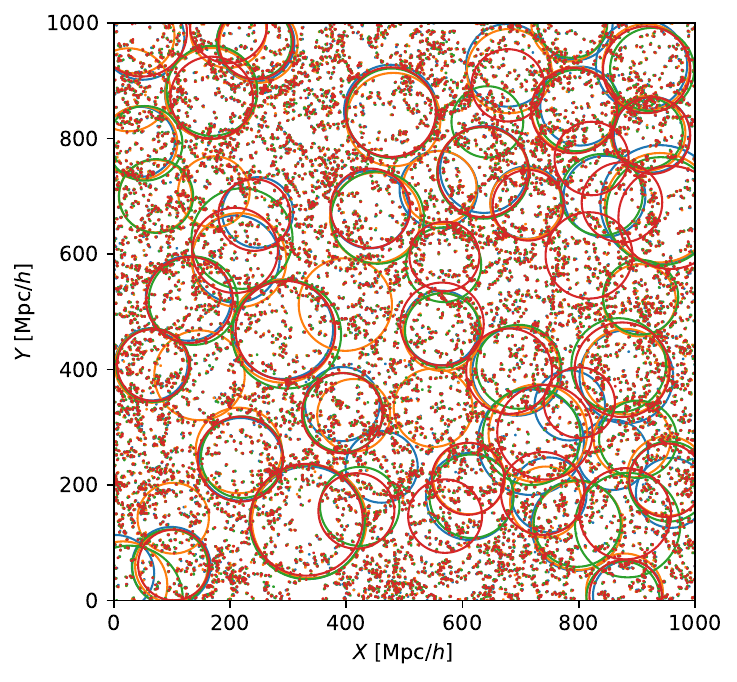}
  \caption{
    Illustration of the instability of deterministic void finders.
    Four times, the same galaxies are randomly jittered by $\sigma=1\,h^{-1}\text{Mpc}$.
    The corresponding VIDE voids are shown as circles (matching in effective radius).
    There are substantial differences between the void catalogs.
  }
  \label{fig:jitter}
\end{figure}
Discarding the probabilistic character of the problem is not only problematic in a formal sense.
It is tied to the observed instability of existing void finders~\citep[e.g.,][]{Sartori2026}.
We demonstrate this instability in Fig.~\ref{fig:jitter} where VIDE void catalogs obtained after small random jitters of the underlying galaxies are displayed.
We see that even though voids (here with minimum radius $60\,h^{-1}\text{Mpc}$) would be expected to decouple from small-scale perturbations, the resulting void catalogs are quite different.
It has also been noticed that computing catalogs over random subsamples of a given galaxy field and averaging their statistics leads to significantly improved cosmological constraining power~\citep{Liu2026}.

One approach towards Bayesian void finding as defined by Eq.~\eqref{eq:catalogp} is through field-level reconstruction of the entire density field where the voids can then be defined~\citep{Leclercq2015a,Leclercq2015b,Malandrino2026}.
While theoretically optimal, the technique is computationally expensive and has to sacrifice small-scale information to be feasible.

Another approach would be a local classification of the density field so as to match existing local algorithms~\citep{Sousbie2013,Cautun2013,Forero-Romero2009,Pomarede2017}.
Indeed, deep learning has been used to learn such classifications~\citep{Kumagai2025,Kololgi2025}.
We see value in catalogs of discrete voids as opposed to point-wise classifications, however.
The field has developed many tools to analyze and theoretically describe such catalogs.
Discrete voids are also convenient for cross-correlations, including with the CMB~\citep{Cai2017,Sartori2025,Granett2008,Melchior2014} and with gamma rays~\citep{Furniss2015,Arcari2022}.

Thus, in this work we develop a general solution to computing samples (i.e.\ void catalogs) from the distribution Eq.~\eqref{eq:catalogp}.
Our solution relies on deep learning, specifically graph neural networks trained with a flow matching objective.
The training is performed in a supervised manner on cosmological simulations.
We emphasize that our method is rather general in the sense that any void definition can be targeted.
That is, the training data can be computed in any way from the available simulations, including from the initial or final density and velocity field.
For any desired void definition motivated by a specific scientific question, our method provides the Bayes-optimal void finder.

After laying out the model design and training procedure in Sec.~\ref{sec:methods}, we will present results through some summary statistics in Sec.~\ref{sec:results}. The discussion in Sec.~\ref{sec:discuss} puts this work in the context of cosmological inference and observations in general machine learning. We conclude in Sec.~\ref{sec:concl}.

\section{Methods}\label{sec:methods}

Mapping the construction of a void catalog to a generative machine learning problem seems to require a somewhat non-standard setup.
We solve the problem by using \emph{test particles} whose coordinates and attributes correspond to the desired void positions and properties.
Indeed, test particles are a known solution for image segmentation problems~\citep[e.g.,][]{Grady2006} and have been used for reconstructions of the cosmic web~\citep{Burchett2020}.
The presentation will be restricted to a cubic box at a fixed redshift with galaxies in real space.
Future work will introduce the (tractable) complications necessary to deal with observational lightcones.
Since the details of the technique are quite technical, we begin with a broad outline in Sec.~\ref{sec:primer} before going into the specifics in Sec.~\ref{sec:train}.
Training data are then described in Secs.~\ref{sec:data} and we conclude the methods by desribing the neural network in Sec.~\ref{sec:arch}.

\subsection{Flow Matching: Primer}\label{sec:primer}

The mentioned test particles are evolved by a neural network to track as closely as possible so-called interpolants.
The method is known as flow matching~\citep{Lipman2023,Liu2022} which is a limit of stochastic interpolants~\citep{Albergo2023,Chen2024}.
Applications of similar methods in cosmology are found in~\citet{Cuesta2024a,Cuesta2024b,Horowitz2025a,Kannan2025}.

In broad terms, flow matching is a particular solution to the problem of describing (conditional) probability distributions with neural networks.
Instead of explicitly learning the target density, the flow operates at the level of individual samples (Eq.~\eqref{eq:sample} below).
We choose a \emph{base distribution} from which samples can be easily generated and then learn a velocity field (Eq.~\eqref{eq:velocity} below) which transforms samples from this base distribution into samples from the desired distribution.
Since only the endpoints of the flow are determined by the base distribution and the target distribution, there is freedom in how the connecting path (interpolant) is chosen.
Our choice of interpolants is given in Eq.~\eqref{eq:interpolant} below.
The velocity field is parameterized through a neural network and learned by randomly pairing samples from the base and target distributions.
The somewhat surprising feature of flow matching is that training on randomly paired samples will yield the correct velocity field at optimum.
Computationally, it is often useful to reduce the randomness (i.e., decrease typical ``distance'' between base and target samples).
We introduce a scheme of reduced randomness matched to our particular problem in the below description.

\subsection{Flow Matching: Details}\label{sec:train}

We introduce a fictitious time $t\in[0, 1]$ which parameterizes the flow between two distributions.
At $t=0$ the test particles follow a trivial base distribution, while at $t=1$ they are supposed to follow the complicated distribution over void catalogs we aim to sample from.

The $\alpha$th test particle is a tuple $v_\alpha(t)$ which contains the position of the particle and further attributes:
\begin{equation}
  v_\alpha(t) = (\vec r_\alpha(t), \vec a_\alpha(t))\,.
  \label{eq:sample}
\end{equation}
For this proof-of-concept work, we take $\vec a_\alpha \in \mathbb{R}^1$ corresponding to the volume of the targeted cosmic void.
There is no limitation on other void properties to predict as part of the $\vec a_\alpha$.

At $t=0$ there are $N_p$ test particles Poisson-initialized in the box at $v_\alpha(0)$.
Time evolution of the test particles is defined as the ODE
\begin{equation}
  \frac{dv_\alpha(t)}{dt} = f_\phi(v_\alpha(t), \{v_{\alpha'}(t)\}_{\alpha'=1 \ldots N_p}, \{g_\beta\}_{\beta=1 \ldots N_g}, t)\,,
  \label{eq:velocity}
\end{equation}
where $g_\beta$ describe the galaxies in which the voids are to be found.
The graph neural network $f_\phi$ is described in more detail in Sec.~\ref{sec:arch}.

During training, fictitious time is sampled uniformly $t\sim[0,1]$ and optimization of the neural network is performed to minimize the objective
\begin{equation}
  \int_0^1 dt \sum_{B \in \substack{\text{training}\\\text{set}}}\,\sum_{\alpha=1 \ldots N_p} \left| f_\phi(v_\alpha(t), \ldots) - \frac{d \hat v_\alpha(t)}{dt} \right|^2\,,
  \label{eq:loss}
\end{equation}
where $B$ is a simulated box and we introduced the \emph{interpolants}
\begin{equation}
  \hat v_\alpha(t) = (1-t) v_\alpha(0) + t \hat v_\alpha(1)\,.
  \label{eq:interpolant}
\end{equation}
Here, $\hat v_\alpha(1)$ corresponds to an individual void contained in the training data.
When trying to match test particles to cosmic voids, two problems arise.

First, we do not know the number of voids $N_v$ a priori, so a one-to-one matching between test particles and voids is impossible.
We solve this by oversampling, i.e.\ using $N_p>N_v$ test particles.
Most voids in a training sample get assigned multiple interpolant indices $\alpha$ such that the total sums to $N_p$.
We randomly set the number of endpoints per void such that in expectation it is proportional to the void's volume and has the minimum variance around it.

Second, given that a typical training sample (simulation box) in our setup contains hundreds of voids, \emph{which} test particle $\alpha$ should be assigned to a given void?
In the original flow matching framework, the assignment would be random.
Such a choice of interpolants would be very difficult to learn, however, since the test particles would need to traverse large distances across the box and this would be unnatural in our locality-biased graph neural network.
We perform the pairing by minimizing the sum of squared Euclidean distances between the initial random test particle positions $\vec r_\alpha(0)$ and the void catalog.\footnote{The described assignment problem is a special case of optimal transport and can be efficiently solved with the Hungarian algorithm~\citep{Kuhn1955}.}
To avoid degeneracy in the minimum-distance assignment, we Gaussian ``jitter'' the void positions in $\hat v_\alpha(1)$ with $0.2\,h^{-1}\text{Mpc}$ standard deviation.

The resulting construction of interpolants is in the spirit of ``data-dependent couplings''~\citep{Albergo2024} but differs from previous literature in that it acts as a permutation, not a continuous transformation.
We point out that the construction might superficially resemble minibatch optimal transport~\citep{Pooladian2023,Tong2024} but is very different due to the couplings between test particles (the existence of minibatch-OT as the zero-coupling limit provides a useful check, however).
Note that the possibility to minimize total traversed path length does not exist in diffusion models.

During inference, we integrate Eq.~\eqref{eq:velocity} according to the Euler rule
\begin{equation}
  v_\alpha(t_{i+1}) = v_\alpha(t_i) + (t_{i+1}-t_i)f_\phi(v_\alpha(t_i), \ldots)
\end{equation}
over the time interval $t\in[0, 1]$.
We find that the resulting paths show somewhat more curvature at the beginning and end, thus opt for the stepping scheme
\begin{equation}
  t_i = 0.9(3\tau_i^2-2\tau_i^3)^{1.5} + 0.1\tau_i\,, m\tau_i = 0 \ldots m-1
  \label{eq:timesteps}
\end{equation}
where we take $m=32$ integration steps.
Our results are not very sensitive to this particular choice of time steps $t_i$.
In Fig.~\ref{fig:trajsteps} in the appendix we demonstrate that the integrated trajectories are very well converged.

As we oversample $N_p > N_v$, the integration of $v_\alpha(1)$ needs to be followed by a procedure to combine test particles into unique voids.
We do this clustering using the friends-of-friends (FoF) algorithm~\citep{Huchra1982,Davis1985} on the $\vec r$-part of the prediction.
Void properties are then computed as the mean of the grouped $v_\alpha(1)$.
The FoF linking length is a hyperparameter but we find little dependence on it for reasonable values.
Thanks to recent developments on differentiable clustering algorithms~\citep{Horowitz2025b} our entire void finder could be made differentiable.
Differentiable pipelines from initial conditions to summary statistics can enable more efficient parameter estimation, for example with Hamiltonian Monte Carlo or raytracing~\citep{Behroozi2025} sampling.

\subsection{Training data}\label{sec:data}

For the purposes of this initial exposition of the deep learning method, we simplify the learning problem to its essential properties.
Thus, we use Quijote simulations~\citep{Villaescusa-Navarro2020} of side length $1\,h^{-1}\text{Gpc}$ run with $512^3$ particles evolving under gravity only.
Galaxies are inserted into the simulations according to a halo-occupation-distribution (HOD) prescription described in Sec.~\ref{sec:data:gals}.
We work at redshift $z=0$ in real space.

For training, we use 12000 simulations from the BSQ set in which 5-parameter $\Lambda$CDM is varied over a wide prior~\citep{Bairagi2025}.
Another 1000 BSQ simulations are used for validation and testing, each.
Each of the BSQ simulations is populated with $~8.6$ randomly drawn HODs on average.
The resulting 115707 galaxy catalogs are further augmented through the usual 48 rotations and transpositions.
These augmentations help the neural network to learn spatial symmetries as it is not inherently equivariant under them.

In addition, we use 15000 simulations at the fiducial point populated with the fiducial HOD parameters for evaluation.

\subsubsection{Galaxies}\label{sec:data:gals}

\begin{table}
  \centering
  \begin{tabular}{lll}
    parameter & prior bounds & fiducial \\
    \hline
    $\Omega_m$ & $[0.1, 0.5]$& $0.3175$ \\
    $\Omega_b$ & $[0.02, 0.08]$ & $0.049$ \\
    $h$ & $[0.5, 0.9]$ & $0.6711$ \\
    $n_s$ & $[0.8, 1.2]$ & $0.9624$ \\
    $\sigma_8$ & $[0.6, 1.0]$ & $0.834$ \\
    \hline
    $\log M_0/M_\text{min}$ & $[0, 1]$ & $0.5$ \\
    $\log M_1/M_0$ & $[0, 3]$ & $1.5$ \\
    $\sigma_{\log M}$ & $[0.1, 0.8]$ & $0.45$ \\
    $\alpha$ & $[0.2, 1.5]$ & $0.85$ \\
  \end{tabular}
  \caption{Prior bounds and fiducial values for the cosmological and  HOD parameters.}
  \label{tab:hod}
\end{table}

We use a standard HOD~\citep{Berlind2002,Cooray2002,Wechsler2018} to populate the gravity-only simulations with galaxies.
The form is taken from~\citet{Zheng2005} and explicitly written out in Appendix~\ref{app:hod}.
HOD parameters are sampled according to uniform priors with bounds given in Table~\ref{tab:hod}.
The HOD fiducial point is taken as the mean of the priors.

Using $M_\text{min}$, we fix the overall number of galaxies per training sample to $N_g = 10^5$, corresponding to a number density of $n = 10^{-4} (\text{Mpc}/h)^{-3}$.
Given the large parameter variations we believe it makes physical sense to constrain the number density.
Furthermore, for this initial work it has computational advantages as each training sample demands about equal computation.
At the Quijote fiducial cosmology and HOD fiducial point, the corresponding $\log M_\text{min}/h^{-1}M_\odot = 13.78$.
We are working with a rather sparse galaxy sample for simplicity in this initial demonstration but do not see any fundamental limitation on galaxy density.\footnote{Compare with $n\sim 10^{-3} (\text{Mpc}/h)^{-3}$ for the final DESI LRG sample. We are also constrained by the 32GB of VRAM on the V100 GPUs we have available.}

\subsubsection{Void definitions}\label{sec:data:voids}

The primary void definition we use in this work are VIDE-voids in the galaxy field~\citep{Sutter2015}.
At first glance, this seems paradoxical: this void definition uses the same information as is available during inference of our graph net flow.
Thus, we are seemingly not taking advantage of our algorithm's Bayesian character at all and simply aiming to replicate a deterministic void finder.
As we shall see, the strong inductive bias in our architecture and training objective lets the network converge to a non-deterministic void finder.
We will argue in Sec.~\ref{sec:discuss} that in some sense the trained network can be seen as \emph{outperforming} its VIDE teacher.

Regardless of these points, it is also beneficial for this initial work to aim for a well-understood baseline performance.
In future work we will learn to predict voids defined on the simulated density and velocity fields and thus take full advantage of our Bayes-optimal method.

The VIDE-catalogs from the galaxy field are truncated at a minimum void radius of $R_\text{min}=60\,h^{-1}\text{Mpc}$.
This value is appropriate for the low tracer density we are working with (mean galaxy separation is $\sim 21.5\,h^{-1}\text{Mpc}$).
Given the range of cosmologies and HODs in our training data, individual boxes contain $N_v \sim 100 \ldots 600$ voids after this cut.
In the following, we denote them \emph{galaxy voids}.

As a test, we also use void catalogs based on the dark matter field.
These voids are found using VIDE run directly on the simulation particles.
We choose this setup so as to test the limits of our method since galaxy voids have little in common with these dark matter-defined ones.
For computational reasons we subsample the simulation particles to $10\,\%$.
Due to the very different tracer density, we cut these catalogs at $R_\text{min}=25.04\,h^{-1}\text{Mpc}$ so as to obtain equal void numbers with the galaxy voids at the fiducial point.
We will see that voids constructed from these high-density tracers are very difficult to identify from the galaxies, introducing large stochasticity and complicating the learning problem.
In the following, we denote them \emph{DM voids}.

We always use VIDE's default volume-weighted barycenter definition for the positions and note that this is a highly nonlocal definition depending on all Voronoi cells belonging to a given void.
One might expect void position definitions based on local density minima to be easier to learn.
Since we are interested to explore the capabilities of our proposed algorithm, we choose the more difficult definition.

\subsection{Architecture and Training}\label{sec:arch}

\begin{figure}
  \centering
  \includegraphics[width=0.8\linewidth]{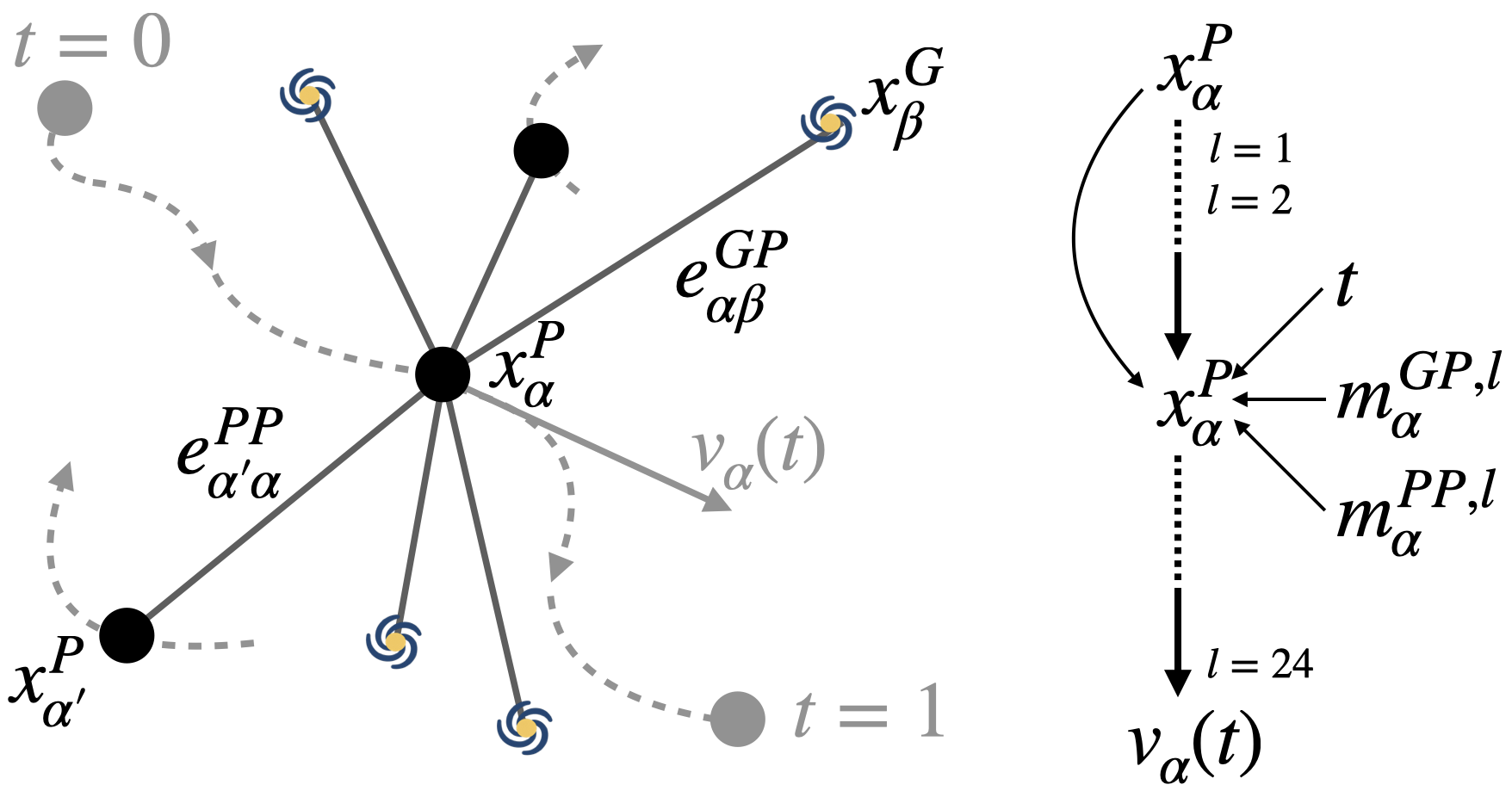}
  \caption{
    Schematic representation of the graph neural network architecture.
    The test particle node features $x^P_\alpha$ get updated through multiple rounds of message passing so as to give the desired velocity $v_\alpha(t)$.
    Messages are computed from both galaxies ($m^{GP}_\alpha$) and other test particles ($m^{PP}_\alpha$) in $k$-nearest neighborhoods.
  }
  \label{fig:arch}
\end{figure}

We construct the test particle fictitious velocity $f_\phi(v_\alpha(t), \{v_{\alpha'}(t)\}_{\alpha'=1 \ldots N_p}, \{g_\beta\}_{\beta=1 \ldots N_g}, t)$ as a graph neural network~\citep{Gori2005,Scarselli2009}.
See Fig.~\ref{fig:arch} for a sketch.
There are two types of nodes, namely test particles $P$ and galaxies $G$.
We also include edges $PP$ and $PG$.
Appendix~\ref{app:feat} describes how we compute the input node and edge features.
We emphasize that the graph topology depends on fictitious time.

The network only updates the test particle features $x^P_\alpha$, since the computational cost of updating the galaxy features $x^G_\beta$ and edges $e^{PG}_{\alpha\beta}$, $e^{PP}_{\alpha\alpha'}$ would be prohibitive.
The network updates $x^P_\alpha$ using message passing~\citep{Gilmer2017,Battaglia2018,Hamilton2018}.
Messages from galaxies to particles are computed in the $l$th round of message passing as
\begin{equation}
  m^{GP,l}_\alpha = \text{Aggr}_{\beta\in\text{knn}(\alpha)}g_l(\{\tilde x^P_\alpha, x^G_\beta, e^{GP}_{\alpha\beta}\})\,,
\end{equation}
where $\text{Aggr}$ is a permutation-invariant aggregation over the $k$-nearest neighbors (in 3-dimensional Euclidean space) and $g_l$ is a multi-layer perceptron (MLP).
Analogous messages $m^{PP}$ are computed from the neighboring test particles.
Although the test particle features $x^P_\alpha$ are updated through subsequent rounds of message passing, we always concatenate them with the initial features to give $\tilde x^P_\alpha$.
The motivation for such skip-connections is to prevent the oversmoothing which can occur in deep neural networks.
We take the aggregation functions as mean and maximum in the ratio 3:1 between different message passing rounds.

Test particle features are updated according to the rule
\begin{equation}
  x^P_\alpha \leftarrow \text{LayerNorm}(x^P_\alpha + \eta \text{FiLM}_t(h_l(\{m^{PP,l}_\alpha, m^{GP,l}_\alpha\})))\,,
\end{equation}
with layer normalization~\citep{Ba2016} for stability, feature-wise linear modulation~\citep{Perez2018} to include conditioning on the fictitious flow time $t$, $\eta$ a learned feature-wise scaling, and $h_l$ another MLP.

Our network architecture does not respect the periodic boundary conditions of the simulation boxes.
This choice is made deliberately looking towards the planned application of the method to observational lightcones.
We also do not construct the architecture in a rotationally or translationally equivariant way, due to the computational cost of respecting these relatively small symmetries.

We use the following numerical settings: $k_{GP}=128$, $k_{PP}=16$ for the $k$-nearest neighbors, MLP width 256 and depth 2, 24 rounds of message passing, for a total of 22 million network parameters.
Activation functions are exponential linear units~\citep[ELU,][]{Clevert2016}.

The scaling of our architecture is naively $O(\lambda N_p^2 + N_p N_g)$ where $\lambda$ depends on hyperparameters.
With the choices adopted in this work the second term dominates.
The given scaling expression assumes a constant survey volume with receptive fields fixed in length units.
It is reasonable to assume that with larger galaxy density there would be more interest in smaller voids and thus the appropriate receptive field could be smaller.
Therefore, the practical scaling might be better than the naive expectation.

\begin{figure}
  \includegraphics[width=\linewidth]{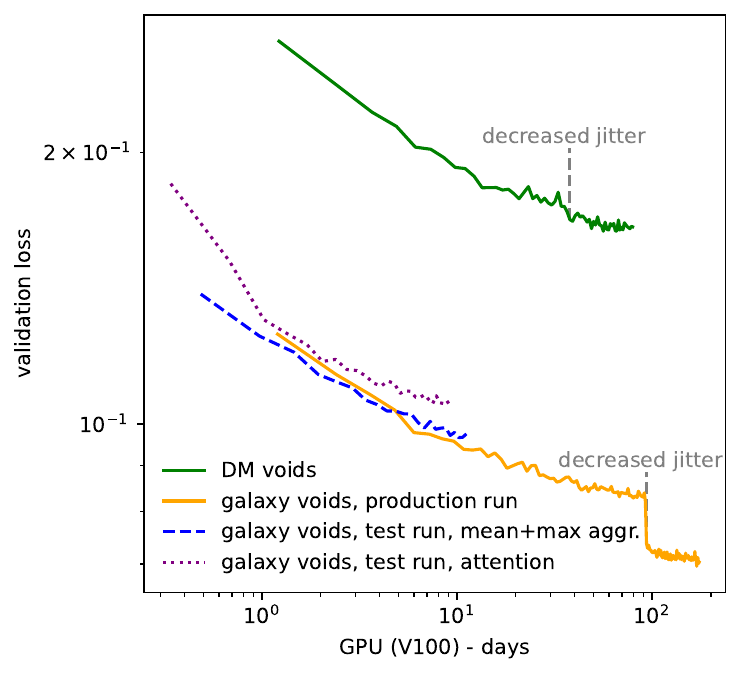}
  \caption{
    Validation loss curves for final production runs (orange, green) and illustrative trial runs with smaller models (blue, purple).
    At the indicated points, the random jittering of target void positions was decreased to the final value.
  }
  \label{fig:loss}
\end{figure}
We implement the network in PyTorch~\citep{Paszke2019} using the PyG library\footnote{\url{https://pytorch-geometric.readthedocs.io}} and optimize it with the Adam algorithm~\citep{Kingma2017}.
Some loss curves are shown in Fig.~\ref{fig:loss}, where we plot the empirical loss on unseen validation data which the model could not have ``memorized'' during training.
The long training runs correspond to the final models examined more closely in Sec.~\ref{sec:results}.
For both galaxy voids (orange) and DM voids (green) we observe typical convergence, albeit to rather different values due to the much larger stochasticity in the latter case.
The galaxy voids case may benefit from further training beyond our computational budget.

Amongst many trial runs, we point out a comparison of two which might be interesting for practitioners.
The blue dashed curve corresponds to a network with similar structure to our final model but smaller scale for faster exploration.
The purple dotted curve is for a model of similar size but with a multi-head attention mechanism~\citep{Velickovic2018} instead of the traditional message aggregation.
It is only computed over the $k$-nearest neighbors regions.
We observe that the traditional aggregation strategy outperforms attention.
We believe that the advantage of traditional architectures over transformer-inspired ones is a rather general phenomenon for cosmological data, the reason being the smoothness of typical fields.
From theoretical works, it is known that transformers exploit sparsity~\citep[e.g.,][]{Mousavi-Hosseini2025}.
Indeed, previous works found analogous advantages of traditional CNN and GNN aggregations over attention~\citep{Hwang2023,Kakadia2025}, although the opposite has been found as well~\citep{Kololgi2025}.
We thus caution against extrapolating results from natural language or everyday images to the rather different problems arising in cosmology.

\section{Results}\label{sec:results}

In this section, we begin by considering the case of galaxy voids.
We remind the reader that the desired mapping is, in principle, deterministic and perfectly learnable.
We first present results computed at the fiducial point (of both cosmology and HOD) and defer parameter variations to the discussion in Sec.~\ref{sec:discuss}.
We use $N_p = 10^3$ test particles and adopt a linking length $b = 15\,h^{-1}\text{Mpc}$ for the FoF step (recall the void size cut $R_\text{min} = 60\,h^{-1}\text{Mpc}$).

\begin{figure}
  \includegraphics[width=\linewidth]{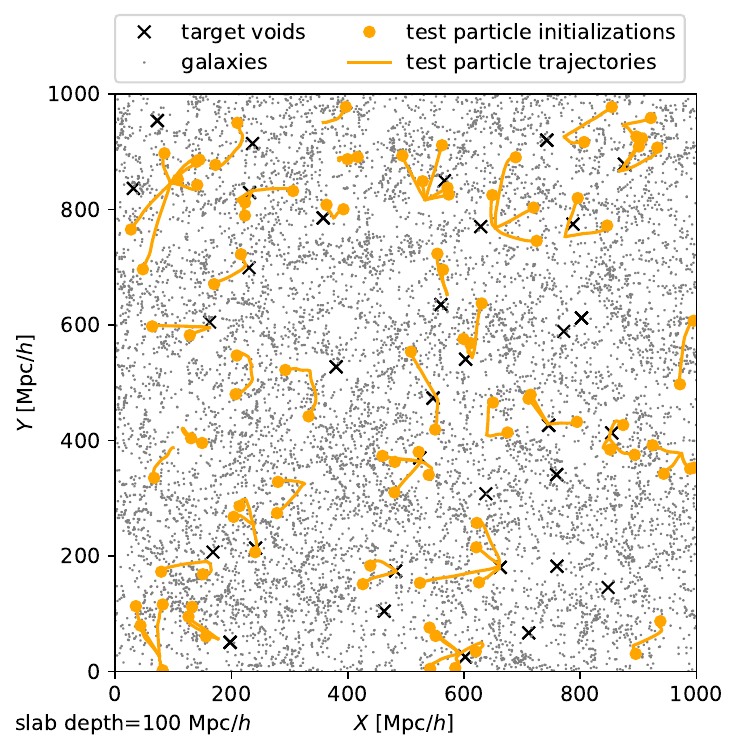}
  \caption{
    Slab of $100\,h^{-1}\text{Mpc}$ thickness in a test sample.
    Grey points are galaxies and black crosses ground-truth VIDE voids.
    Test particles are randomly initialized in the filled orange circles and then evolved by the graph network along the orange trajectories.
    The test particles also carry a prediction of void volume which is not visualized.
    See Fig.~\ref{fig:trajsamples} in the appendix for multiple random initializations of the test particles.
  }
  \label{fig:traj}
\end{figure}
In Fig.~\ref{fig:traj} we begin by presenting a visual impression of the trained void finder's performance.
We show a slab of $100\,h^{-1}\text{Mpc}$ thickness in one of the test samples.
Galaxies are shown as grey points, the target void centers (computed by VIDE) as black crosses, and the test particle trajectories as orange lines.
The randomly sampled starting points of these trajectories are marked as filled orange circles.
The trajectories have small curvatures.
Further, whenever multiple test particles converge because the model deems them to belong to a single void, the convergence is very good.
These observations indicate a well-trained model.
In many instances there is a clear correspondence between trajectory ends and ground-truth voids.
This is, however, not universally the case.
Thus, the trained void finder behaves differently than the mapping it was trained on.

\begin{figure}
  \includegraphics[width=\linewidth]{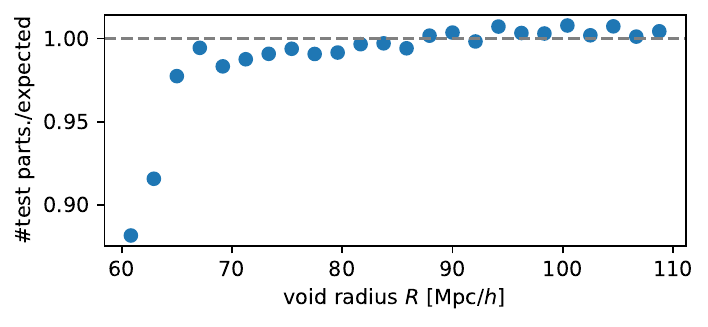}
  \caption{
    Check for the consistency condition that, in expectation, number of converging test particles should be proportional to void volume.
  }
  \label{fig:cnt}
\end{figure}
A useful check is how many test particles end up being grouped together in voids of various sizes.
According to our construction of the training set, we expect the number of test particles converging towards a single void to be proportional to the void volume.
In Fig.~\ref{fig:cnt}, we show how the model behaves relative to this expectation.
For almost all void sizes the empirical behavior is correct, except for the smallest voids for which a $\sim 10\,\%$ discrepancy appears.
Our interpretation is that for a small set of ``failed'' test particles, the model is unable to find a position towards which it should evolve them.
The adopted behavior in these cases seems to be to evolve the position only very little and to drive the volume prediction towards the lower end of the prior.
Due to the decreasing void size function, this could be Bayes-optimal given the model's limitations.

\begin{figure}
  \includegraphics[width=\linewidth]{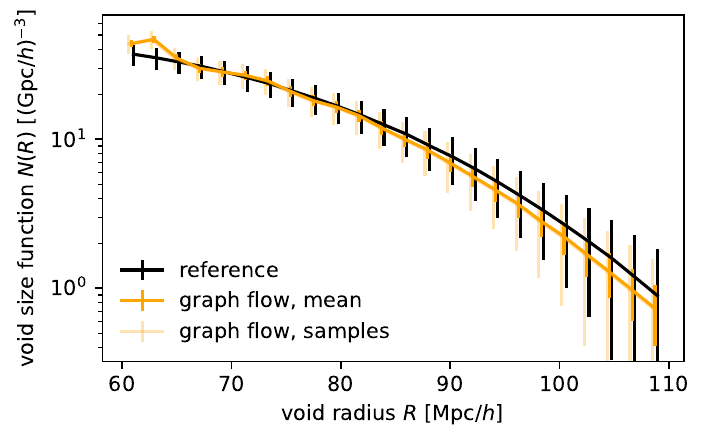}
  \includegraphics[width=\linewidth]{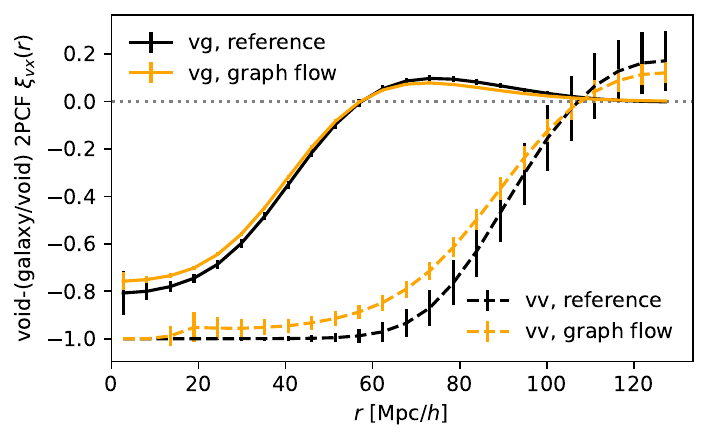}
  \caption{
    Summary statistics of galaxy voids at the fiducial cosmology and HOD.
    Black is VIDE ground truth, orange our model's prediction.
    \emph{Top panel:} void size function. \emph{Bottom panel:} void-galaxy and void-void correlation functions.
  }
  \label{fig:stats}
\end{figure}
In Fig.~\ref{fig:stats} we compare the main summary statistics between model predictions (orange) and ground-truth VIDE-voids (black).

The void size function (VSF) in the top panel matches relatively well except for an overprediction at small radii and a slight underprediction at large radii.
The overprediction is consistent with the previously developed picture of ``failed'' test particles.
The underprediction can likely be understood as convergence to the mean. 
For the void size function, we also show the statistical scatter as error bars.
Light orange is the scatter including model stochasticity (i.e.\ the model only evolves a single population of test particles per test sample),
while dark orange is the sample variance alone (i.e.\ the model is averaged over multiple draws of test particles).
We observe that the trained model has substantial stochasticity but ends up matching the scatter in the VIDE VSF quite closely.
When the model's stochasticity is averaged over, however, the VSF is substantially \emph{less} noisy than its teacher.
We will discuss this more in Sec.~\ref{sec:discuss}.

The void-galaxy correlation function (i.e.\ average void profile) and void-void correlation function\footnote{Correlation functions are computed with pycorr, \url{https://py2pcf.readthedocs.io}.} in the bottom panel are consistent with our previously developed picture of a well-trained model with a small population of ``failed'' test particles.
Both statistics match the ground truth relatively well while indicating slightly less underdense and slightly less mutually excluding voids.
In the void-void correlation function we can clearly see a feature matching the FoF linking length $b = 15\,h^{-1}\text{Mpc}$.

\begin{figure}
  \includegraphics[width=\linewidth]{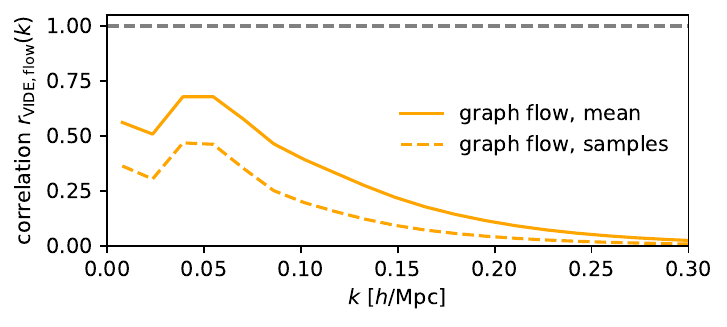}
  \caption{
    Cross-correlation between our model's output void catalog and the VIDE ground truth.
    Note that we have not subtracted any shot noise from the auto power spectra (it is not Poissonian).
    Thus, we are underestimating $r(k)$.
  }
  \label{fig:pkcr}
\end{figure}
A generative model can sometimes reproduce summary statistics quite well without actually solving the intended task.
Therefore, in Fig.~\ref{fig:pkcr} we show the cross-correlation between our generated void catalogs and the reference ones.
Note that as macroscopic objects voids have unknown non-Poissonian stochasticity and thus we cannot subtract shot noise from the auto power spectra.
Therefore, $r(k)$ is systematically underestimated.
The mean of the generated field yields higher $r(k)$ than the samples, consistent with lower stochasticity.
The correlation decreases beyond $k\sim 0.05\,h\text{Mpc}^{-1}$, roughly corresponding to the considered void size of $R > 60\,h^{-1}\text{Mpc}$.
While difficult to interpret, we conclude that our generated void catalogs are generally correlated with the reference ones, in agreement with the qualitative observations in Fig.~\ref{fig:traj}.

\subsection{DM voids}\label{sec:dmvoids}

After the main results on galaxy voids, we now consider the more difficult problem of finding voids as identified by VIDE in the (subsampled) simulation particle field.
Since for these DM voids we choose the size cut $R_\text{min} = 25.04\,h^{-1}\text{Mpc}$, we adopt a linking length $b = 5\,h^{-1}\text{Mpc}$.

\begin{figure}
  \includegraphics[width=\linewidth]{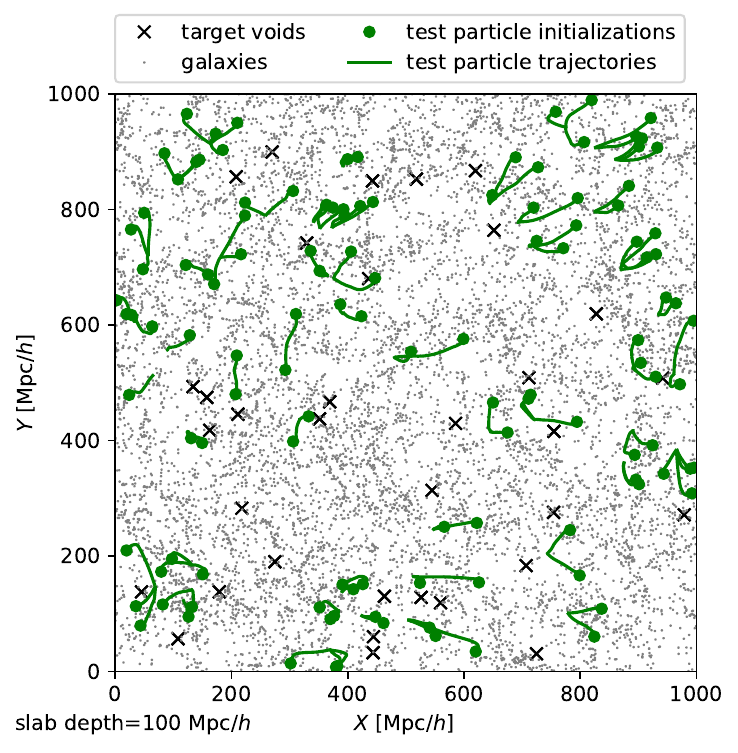}
  \caption{
    As Fig.~\ref{fig:traj} but for DM voids.
  }
  \label{fig:dmtraj}
\end{figure}
Fig.~\ref{fig:dmtraj} shows example trajectories generated by the trained model, analogous to Fig.~\ref{fig:traj} for the galaxy voids case.
We observe little correspondence between the trajectory endpoints and the ground truth voids.
Trajectories are generally more curved as compared to the galaxy voids model.
These observations agree with the intuition that the mapping between galaxies and DM voids has large stochasticity.

\begin{figure}
  \includegraphics[width=\linewidth]{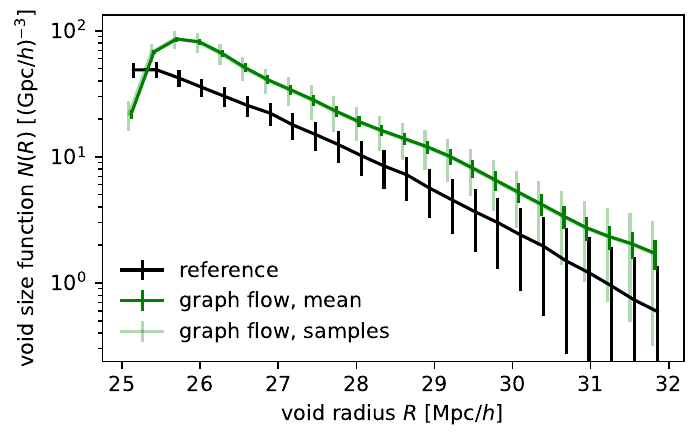}
  \includegraphics[width=\linewidth]{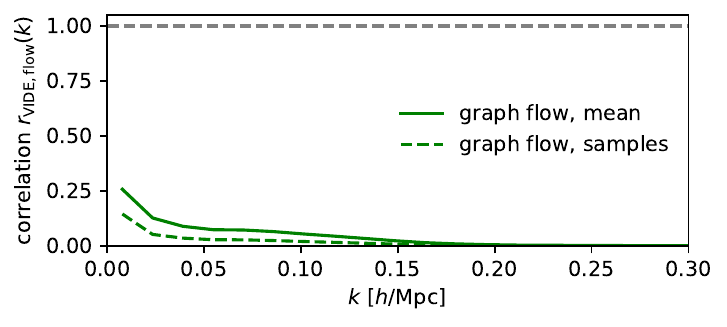}
  \caption{
    \emph{Top panel:} Comparison of void size function for DM voids (c.f.\ Fig.~\ref{fig:stats}).
    \emph{Bottom panel:} Cross-correlation for DM voids (c.f.\ Fig.~\ref{fig:pkcr}).
  }
  \label{fig:dmstats}
\end{figure}
More quantitatively, the top panel in Fig.~\ref{fig:dmstats} shows the DM VSF.
The model generally overpredicts the VSF except for the smallest voids.
This seems connected to test particles not converging as much as in the galaxy voids case.
The picture is consistent with our previous observation that faced with large uncertainty the model produces a population of ``failed'' trajectories that do not converge with other particles.
The bottom panel in Fig.~\ref{fig:dmstats} shows that the cross-correlation with VIDE's output is expectedly lower than in the case of galaxy voids,
but importantly there is still non-zero correlation which means that the model has learnt something meaningful instead of just randomly reproducing summary statistics. 

It is well-known that the void catalogs returned by traditional algorithms are highly dependent on tracer density~\citep{Schmidt2001,Sutter2014,Schuster2023}.
Thus, it is not surprising that supplied only with sparse galaxy catalogs our algorithm struggles to find the DM voids.
Our results are consistent with the hypothesis that the network learned the Bayes-optimal mapping under limited information.

\section{Discussion}\label{sec:discuss}

\begin{figure}
  \includegraphics[width=\linewidth]{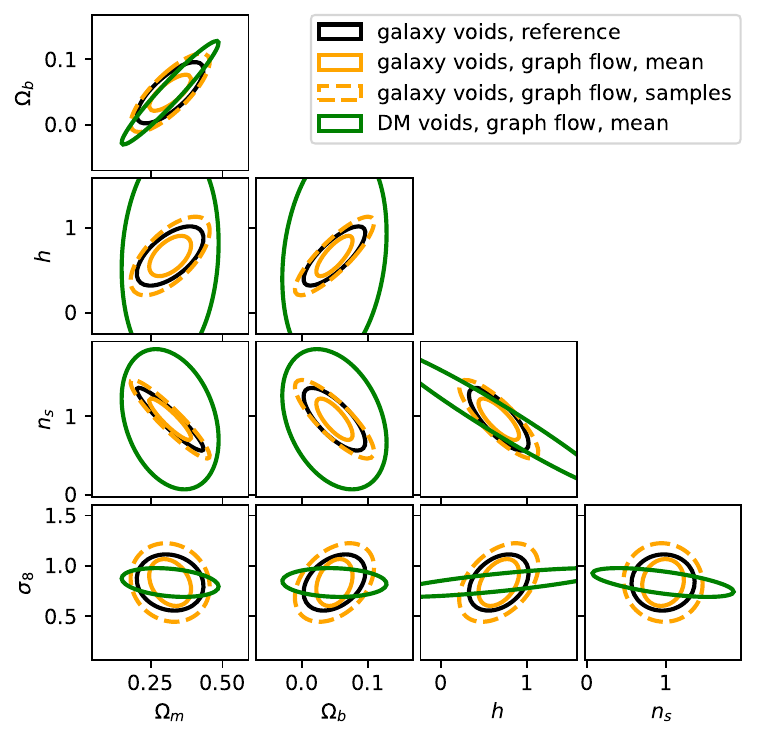}
  \caption{
    Visualization of the cosmological information in void size function and void-galaxy correlation function.
    Marginalization over HOD parameters is included.
  }
  \label{fig:fisher}
\end{figure}
After developing intuition for the kind of void finding algorithm our models have learned, we now turn to discuss usability and interpretation.
A typical use for a void catalog would be to constrain cosmological parameters using summary statistics.
We consider the combined data vector of void size function and void-galaxy correlation function.
(The void-void correlation function has very low signal-to-noise.)
The Fisher information, after marginalizing over HOD parameters, is shown in Fig.~\ref{fig:fisher}.\footnote{Numerical derivatives are computed using the 1000 Quijote BSQ simulations we set aside for testing. The derivatives are approximated by linearizing the entire parameter space. Restricting to a smaller ball around the fiducial point yields some changes in the Fisher matrix but a consistent picture.}
We emphasize that the figure is to be understood not as a forecast but rather as a convenient visualization of \emph{local} signal-to-noise.

Focusing on the galaxy voids (black and orange contours) first, we observe that the orientation of ellipses is quite consistent between the VIDE result and our graph flow.
Furthermore, the information from the mean of summary statistics in graph flow catalogs is higher than the one from VIDE's deterministic catalogs.
Thus, the model's outputs respond correctly to cosmological parameters; no significant convergence towards the training set's mean is observed.

We also show the result from DM voids as found by our graph flow in green.
Direct comparison of contours is difficult but we believe that by cutting catalogs to the same size at the fiducial point (using $R_\text{min}$) it is still meaningful.
Statistics of DM voids are competitive only for $\sigma_8$.
We thus conclude that more work is needed to find a void definition that is learnable by our graph flow and can substantially improve cosmological constraining power.

Going back to the galaxy voids, how do we interpret the fact that our graph flow has learned a non-deterministic void finder which in cosmological constraining power \emph{outperforms} its teacher?
Any finite-size model trained with finite resources will be biased away from the deterministic point defined by the VIDE teacher.
The training algorithm introduces further biasing towards minima with good generalization properties.
We therefore interpret our trained graph flow as a \emph{regularized VIDE}.
The exact regularization is of course impossible to write down but it seems to be useful as indicated by the cosmological information.
Indeed, similar cases of a regularized student outperforming its teacher are known in the machine learning literature~\citep{Xie2020,Furlanello2018}.

\section{Conclusions}\label{sec:concl}

We have developed a Bayes-optimal method to probabilistically sample void catalogs conditioned on observed galaxies.
The targeted void definition is arbitrary and therefore can rely on information that is only available in simulations.
Our algorithm relies on a deep graph neural network that predicts fictitious velocities of test particles to match a defined flow.
Through summary statistics and cosmological information we have demonstrated that our learned flow does indeed predict useful void catalogs.

In practice, the developed algorithm is conceptually easy to apply in a broad range of applications.
Given a galaxy catalog, it produces an arbitrary number of independently sampled void catalogs.
Most existing analysis pipelines that currently work with deterministic void finders will be able to directly adopt our new stochastic void finder.
Measurements need to be run for each void catalog and the final summary statistics averaged over.

Having developed the initial framework in an idealized situation (simulation box, real space, low tracer density, etc.), we will incorporate practical complications in future work.
We aim to exploit the method for a practically usable void finder.
A central question will be which target void definition should be adopted.
We found that naively targeting VIDE voids in the simulation particles yields a mapping dominated by stochasticity.
The DM voids our network is able to identify appear to contain less cosmological information than galaxy voids.
Thus, we plan to identify ``middle ground'' definitions which can take advantage of our Bayesian method while still having high mutual information with sparse observed galaxies.
We see promise in definitions that use the simulated velocity field.

On the machine learning side, the simple flow matching setup works well but could be improved by adding stochastic terms or replacing the ODE with a diffusive SDE.
There might be room for reinforcement learning to improve the output void catalogs beyond the simple FoF clustering step we have been using.
It will also be interesting to explore more uniform base densities, for example glassy initial distributions for the test particles instead of the current Poisson one.

The method of test particles evolving under a conditioned flow is much more general than void finding.
Application to filament finding is underway.
Another useful application could be insertion of halos or galaxies into low-resolution cosmological simulations, similar to, e.g.,~\cite{Pandey2024,Pandey2025a,Pandey2025b}.

Extracting \emph{robust} information from cosmological data is more difficult than signal-to-noise maximization.
Indeed, direct application of deep learning to field-level data often leads to severe biases even in simple situations~\citep[e.g.,][]{Bayer2025}.
We believe that guard-railing the extracted information with understood physical concepts is a promising avenue towards integrating the power of machine learning into cosmological inference.

\section*{Acknowledgements}

I thank Alice Pisani, Pierre Boccard, and Ben Horowitz for reading an earlier version of this manuscript and giving extremely helpful comments.
I thank the organizers and participants of the Voids@CPPM 2025 workshop, in particular Alice Pisani, Nico Schuster, and Pierre Boccard.
I thank Yu Liu and Cheng Zhao for sharing their work and discussions.
I thank Carolina Cuesta-Lazaro, Giovanni Verza, Peter Behroozi, Xavier Prochaska, and Jia Liu for useful discussions.
I thank the anonymous referee for constructive comments which helped improve the manuscript.
LT is supported by JSPS KAKENHI Grant 24K22878.
The Kavli IPMU is supported by World Premier International Research Center Initiative (WPI), MEXT, Japan.

\bibliography{main}

\appendix

\section{Halo Occupation Distribution}\label{app:hod}

The central galaxy is placed at the halo center with probability
\begin{equation}
\overline N_\text{cen} = \frac{1}{2}\left[
  1 + \text{erf}\left( \frac{\log M - \log M_\text{min}}{\sigma_{\log M}} \right)
  \right]\,,
\end{equation}
and the number of satellites is drawn from a Poisson distribution with mean
\begin{equation}
\overline N_\text{sat}  = \overline N_\text{cen}
  \left( \frac{M - M_0}{M_1} \right)^\alpha\,.
\end{equation}
The satellites are distributed isotropically according to an NFW profile~\citep{Navarro1997}
and the concentration model of~\citet{Duffy2008},
using the analytic solution for the inverse NFW CDF from~\citet{Robotham2018}.

\section{Feature engineering}\label{app:feat}

The edge features are taken as
\begin{equation}
  e^{GP}_{\alpha\beta} = \{ \Delta \vec r_{\alpha\beta}, |\Delta \vec r_{\alpha\beta}| \}\,,
\end{equation}
and analogous for $e^{PP}_{\alpha\alpha'}$.

For the node features we aim to approximately respect translational symmetry by projecting the coordinates on a Fourier basis.
Specifically, we compute (in the natural interpretation of the notation)
\begin{equation}
  x^P_{\alpha,r} = \{ \sin, \cos \}(L^{-1}\pi\nu\vec r^P_\alpha)\,,\nu = 2^{0 \ldots m-1}\,,
\end{equation}
where $L=1\,h^{-1}\text{Gpc}$ is the box size and $m=14$.
The transformation
\begin{equation}
  x^P_{\alpha,r} \leftarrow \text{erf}^{-1}(2\pi^{-1}\arcsin x^P_{\alpha,r})
\end{equation}
Gaussianizes these Fourier features.
Furthermore, we compute a number of summary statistics of the local particle distribution.
This is particularly important for the galaxy nodes as these do not receive updates in message passing.
Using the set
\begin{equation}
  d_\alpha = \{ \Delta \vec r_{\alpha\beta} \}_{\beta \in \text{knn}(\alpha)}
\end{equation}
and the corresponding sets $s_\alpha$ of vector lengths and $u_\alpha$ of unit vectors, we compute the density proxies
\begin{equation}
  x^P_{\alpha,\delta} = \{ \text{max}, \text{mean}, \text{std} \}(s_\alpha)\,,
\end{equation}
the gradient proxies
\begin{equation}
  x^P_{\alpha,\nabla} = \sum u_\alpha\,,
\end{equation}
and from the mass matrix
\begin{equation}
  \sum d_\alpha \otimes d_\alpha
\end{equation}
the eigenvalues $x^P_{\alpha,\lambda}$ and major axis $x^P_{\alpha,v}$.
The full feature vector
\begin{equation}
  x^P_\alpha = \{ x^P_{\alpha,r}, x^P_{\alpha,\delta}, x^P_{\alpha,\nabla}, x^P_{\alpha,\lambda}, x^P_{\alpha,v} \}
\end{equation}
has dimension 96.
The construction of galaxy node features $x^G_\beta$ is analogous.
We whiten all node and edge features to zero mean and unit variance.

\newpage
\section{Example test particle trajectories}

\begin{figure}
  \includegraphics[width=0.24\linewidth]{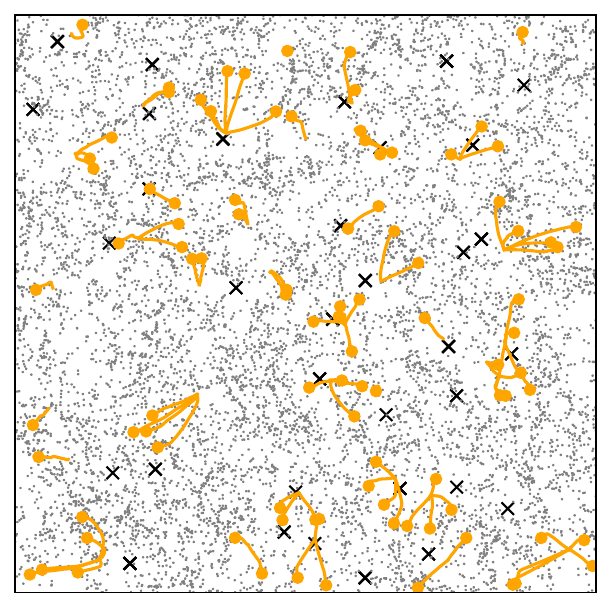}
  \includegraphics[width=0.24\linewidth]{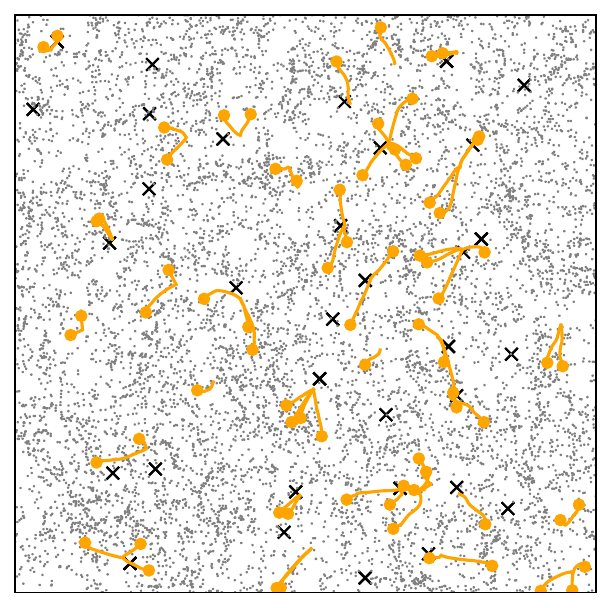}
  \includegraphics[width=0.24\linewidth]{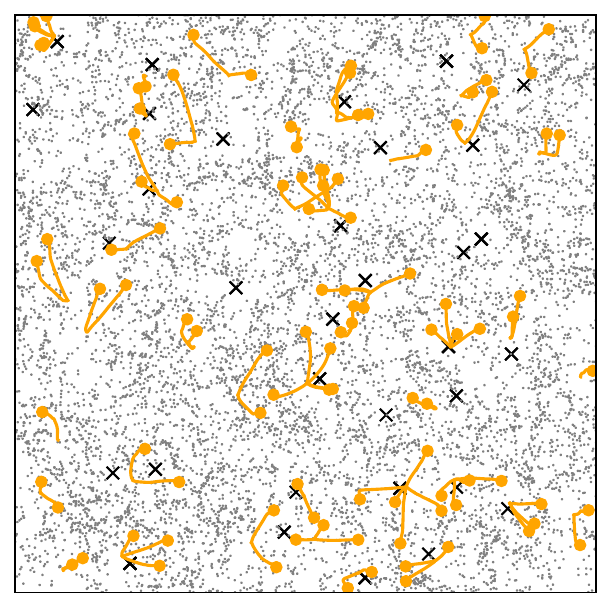}
  \includegraphics[width=0.24\linewidth]{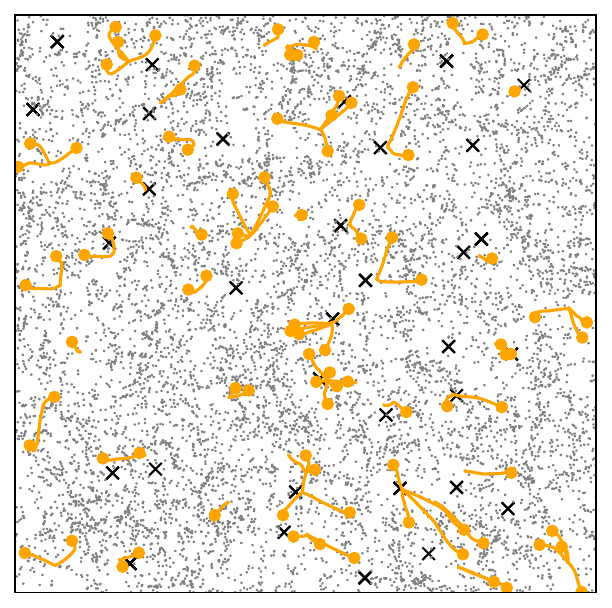}
  \caption{
    Variation of trajectories with test particle initial conditions, compare to Fig.~\ref{fig:traj}.
    Each panel is run with identical galaxies but varying random seed for the test particle initial conditions.
    We observe substantial variation in the trajectory endpoints.
  }
  \label{fig:trajsamples}
\end{figure}
In order to develop further intuition for the variation of output void catalogs with different initial conditions, and thus the level of stochasticity learned by our network, in Fig.~\ref{fig:trajsamples} we show analogs of Fig.~\ref{fig:traj} when the test particle initial conditions are varied.

\begin{figure}
  \includegraphics[width=0.24\linewidth]{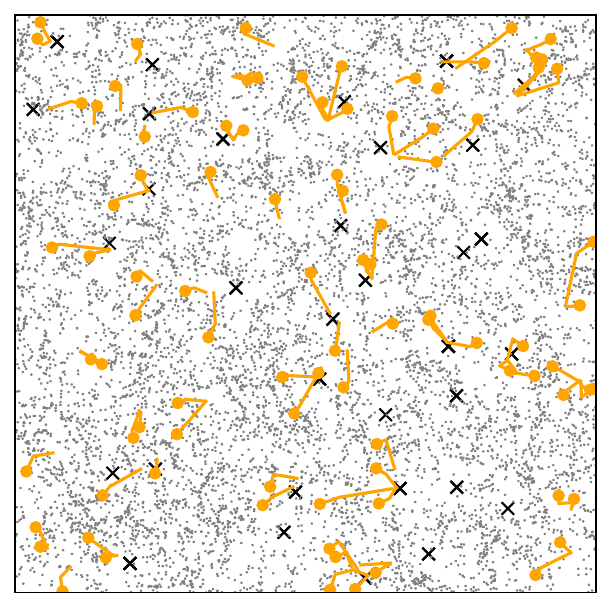}
  \includegraphics[width=0.24\linewidth]{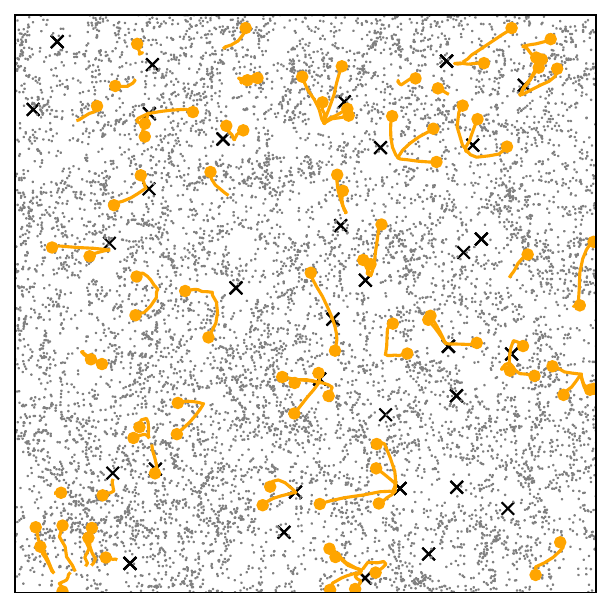}
  \includegraphics[width=0.24\linewidth]{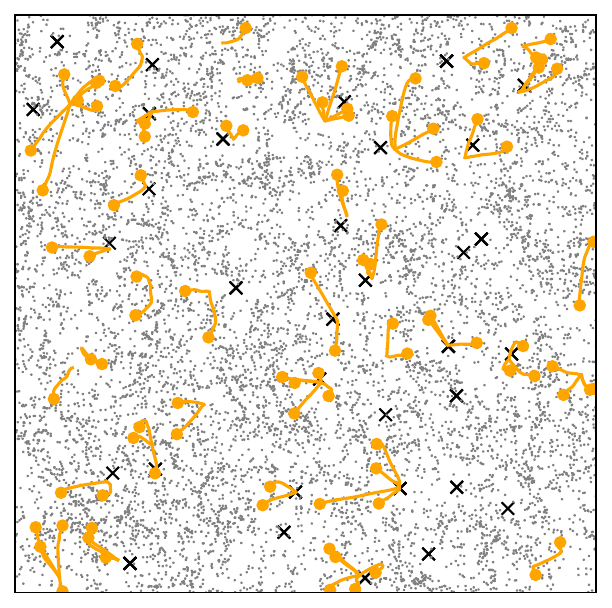}
  \includegraphics[width=0.24\linewidth]{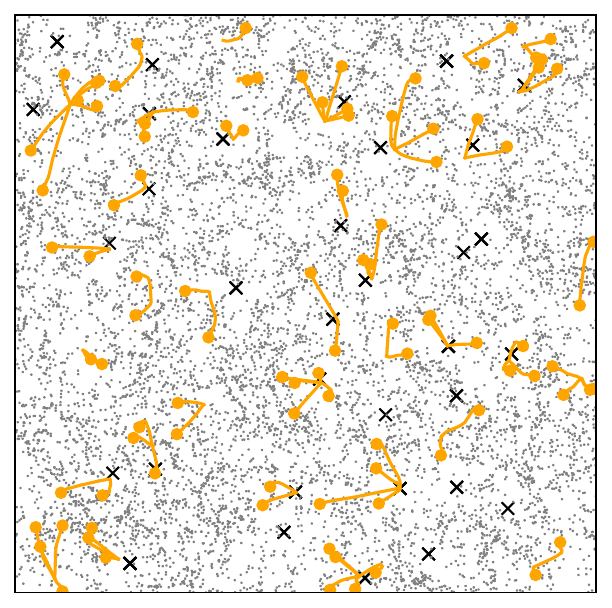}
  \caption{
    Variation of trajectories with integration accuracy, compare to Fig.~\ref{fig:traj}.
    From left to right, we use 2, 8, 32 (fiducial choice), and 128 time steps in the integration of the flow ODE Eq.~\eqref{eq:velocity} while keeping the random seed fixed.
    The convergence with number of integration steps is good.
  }
  \label{fig:trajsteps}
\end{figure}
In Fig.~\ref{fig:trajsteps} we check for convergence of test particle trajectories with respect to the number of integration steps.
The fiducial choice of 32 steps adopted in the main text is seen to be extremely well converged.
This is a consequence of the network's being well-trained and thus predicting low-curvature trajectories.

\end{document}